\documentstyle[12pt]{article}
\begin{document}

\title {Gauge Coupling Unification in GUT and String Models
\thanks{Supported in part by the Polish Committee for Scientific
        Research and EC under the contract ERBCIP-DCT94-0034}}

\author{Piotr H. Chankowski,
Zbigniew P\l uciennik\\
Institute of Theoretical Physics, Warsaw University\\
ul. Ho\.za 69, 00--681 Warsaw, Poland.\\
\\
Stefan Pokorski
 \thanks{On leave of absence from Institute of Theoretical
Physics, Warsaw University}
\\
Max--Planck--Institut f\"ur Physik\\
Werner--Heisenberg--Institut\\
F\"ohringer Ring 6, 80805 Munich, Germany\\
\\
and\\
Costas E.  Vayonakis
\thanks{On leave of absence from Physics Department,
 University of Ioannina, GR--451 10 Ioannina, Greece}
\\
School of Mathematical and Physical Sciences\\
 University of Sussex,\\
Brighton BN1 9QH, U.K.
}
\maketitle
\vspace{2cm}
\vspace{-18.5cm}
\begin{flushright}
{\bf IFT--95/9 \\
     MPI--Pth/95-58 \\
     SUSX-TH-95/17
}
\end{flushright}
\vspace{15.0cm}

\begin{abstract}
The results for the running of the gauge couplings in the MSSM are
up-dated by proper inclusion of all low scale effects. They are
presented as predictions for the strong coupling constant in the
 scenario with only two parameters at the GUT scale ~
($\alpha_{U}$ ~and ~$M_{U}$) ~and as a mismatch of the couplings at
the scales ~$\sim 3 \times 10^{16}$ GeV ~and ~$4\times10^{17}$ GeV,
 ~when
all three couplings are taken as the experimental input.
\end{abstract}

\newpage
{\bf 1.} The gauge coupling unification \cite{su5} within the
Minimal Supersymmetric Standard Model (MSSM) \cite{dim} has been widely
publicized as a successful prediction of SUSY--GUTs [2--6]. It is also often
discussed in the context of stringy unification, with ~
$M_{ST}\simeq4\times10^{17}$ GeV ~\cite{ib2}. In this paper we up-date the
results for the running of the gauge couplings in the MSSM by proper inclusion
of all low energy effects such as the best precision of the input
parameters at the electroweak scale and the non--logarithmic contribution from
the superpartner thresholds \cite{FAGRI,unif,barbieri,bagger}.

The unification idea is predictive with respect to the behaviour
of the ~$SU(3)\times SU(2)\times U(1)$ ~gauge couplings
if physics at the GUT scale can be described in terms of only
two parameters: ~$\alpha_{U}$ ~and ~$M_{U}$ ~(minimal unification).
Then we can predict e.g. ~$\alpha_s(M_Z)$ ~in terms of ~
$\alpha_{EM}(M_Z)$ ~and ~$\sin^2\theta_W(M_Z)$ ~
(it is worth remembering that ~$\sin^2\theta_W(M_Z)$ ~
and ~$\alpha_s(M_Z)$ ~are at present known with 0.1\% and 10\%
accuracy, respectively). More precisely, the prediction for the
strong coupling constant in addition depends on the superpartner
spectrum which  will, hopefully, be known from experiment. For
now, these are free parameters  and, denoting them globally by ~$T_{SUSY}$ ~
(see the discussion in section 2) we get
\begin{eqnarray}
\alpha_s(M_Z) &=& F(\sin^2 \theta_W(M_Z),\alpha_{EM}(M_Z),T_{SUSY})
\end{eqnarray}

This approach may, however, be too restrictive as it is generally
expected that there are non-negligible GUT/string threshold corrections
to the running of the couplings  (such as heavy threshold and higher
dimension operator effects). Then, strictly speaking, all predictivity is
lost. However, it is still very interesting to reverse the problem:
take the values of all the three couplings at ~$M_Z$ ~as input
and use the bottom-up approach to study the convergence of the couplings
in the framework of the MSSM. With the same precision calculation
and as a function of the SUSY spectrum one can, then, discuss the
mismatch of the couplings at any scale of interest and for any value
of ~$\alpha_s(M_Z)$, ~within its 10\% experimental uncertainty.
It is convenient to introduce the ``mismatch'' parameters at scale ~$Q$:
\begin{eqnarray}
D_i(Q)= {{\alpha_i(Q)-\alpha_2(Q)} \over \alpha_2(Q) }
\label{eqn:di}
\end{eqnarray}
and
\begin{eqnarray}
\Delta_i(Q)= {1 \over \alpha_i(Q)}- {1 \over \alpha_2(Q)}
\label{eqn:deltai}
\end{eqnarray}
(the latter are directly related to large scale threshold corrections).
Of particular interest are ~$D_3(M_{U})$, ~where ~$M_{U}$ ~is
defined as the scale of unification of the ~$SU(2)\times U(1)$ ~
couplings (i.e. the scale at which ~$D_1=\Delta_1=0$), ~and ~$D_i(M_{ST}), ~
i=1,3$, ~with ~$M_{ST}=4\times 10^{17}$ ~GeV \cite{kaplunovsky,ib2}
(and corresponding ~$\Delta_i$s). ~Clearly, we get this way constraints
on physics at the high scale, if it is supposed to have
 unification and
the MSSM as the low energy effective theory. We can also read this information
as a hint whether the latter two assumptions look plausible.

In this paper we present our results both as the prediction
for ~$\alpha_s(M_Z)$ ~in the minimal unification scenario
and as a prediction for the mismatch parameters at ~$M_{U}$ ~
and ~$M_{ST}$, ~as a function of ~$\alpha_s$. ~
\vskip 0.5cm

{\bf 2.} We begin with the discussion of the experimental information.
Let us first suppose that the (non-supersymmetric) SM is the correct
effective theory at the electroweak scale. In this theory the couplings ~
$g_3, ~g_2, ~g_1$~ of the ~$SU(3)\times SU(2)\times SU(1)$ ~gauge groups
, at $M_Z$~ and in the $\overline{MS}$~scheme are usually quoted
as the values of $\alpha_{EM}(M_Z),~\sin^2\theta_W(M_Z)$~and $\alpha_s(M_Z)$.
The electromagnetic coupling constant and the Weinberg angle in
the SM are now known with very high precision. The value of
$\alpha_{EM}(M_Z)$~ is obtained from the on-shell
        $\alpha_{EM}^{OS}=1/137.0359895(61)$
via the 1-loop RG improved relation
\cite{SIRLIN}:
\begin{eqnarray}
\alpha_{EM}(M_Z) = {\alpha^{OS}_{EM}\over 1-\Delta\hat\alpha}
\label{eqn:alhat}
\end{eqnarray}
where
\begin{eqnarray}
\Delta\hat\alpha = 0.0682\pm0.0007
+{7\alpha\over2\pi}\log{M_W\over M_Z}
-{8\alpha\over9\pi}\log{m_t\over M_Z}
\label{eqn:dalfa}
\end{eqnarray}
The main uncertainty comes from the continuous  hadronic
contribution to the photon propagator. We explicitly show
the top quark mass dependence of ~$\alpha_{EM}(M_Z)$. ~

The most precise value of ~$\sin^2\theta_W(M_Z)$ ~in the ~$\overline{MS}$ ~
scheme is at present obtained in terms of ~$G_F$, ~$M_Z$ ~and ~
$\alpha_{EM}$. ~The result depends on ~$m_t$ ~and ~$M_{\phi^0}$ ~(the top
and the SM Higgs boson masses respectively) and to a very high precision is
given by the following effective formula
\footnote{
Notice the small change as compared to ref. \cite{unif} which is due
to the inclusion of QCD corrections according to ref. \cite{KNIEHL}.
Another, frequently used, fit is given in
ref. \cite{LP1} and its update in \cite{LP2}.}
\cite{unif}:
\begin{eqnarray}
\sin^2\theta_W(M_Z)= 0.23166 \pm 0.0003  &+& 5.4\times10^{-6} h
                        - 2.4\times10^{-8} h^2\nonumber\\
                        &-& 3.03\times10^{-5} t
                        - 8.4\times10^{-8} t^2
\label{eqn:fits}
\end{eqnarray}
where ~$h\equiv M_{\phi^0}-100$ ~and ~$t\equiv m_t - 165$ ~(both masses
in GeV). The main source of the error is again the hadronic uncertainty
in the photon propagator. E.g. for ~$m_t=180$ GeV ~and ~$M_{\phi^0}=100$
GeV ~we get ~$\sin^2\theta_W(M_Z)=0.2312$.

The value of ~$\alpha_s(M_Z)$~ is known with much worse precision and
depending on the method of determination, the values in the range 0.11-0.13
are quoted \cite{hinchliffe}. It is interesting that the lower part
of this range is favoured by low energy determinations of ~$\alpha_s$ ~
\cite{bethke,shifman} and by a fit to all electroweak data in the framework
of the MSSM \cite{fit}.

Once ~$g_i$s ~at ~$M_Z$ ~in the SM are extracted from the data,
the 2-loop RGE can be used to get them at higher scales.
Passing through the thresholds of superpartners the running
of the couplings is subject to subsequent modifications of
 the ~$\beta$--functions \cite{RR} with,
finally, MSSM RG equations above  all  the thresholds.
Treating the threshold corrections at the 1-loop level
(consistently  with the 2-loop RGE) this procedure gives
\footnote{$\alpha_3^{SM}(M_Z)$ ~differs from ~$\alpha_s(M_Z)$~ by
a threshold correction from the top quark: ~
$\alpha^{-1}_3=\alpha^{-1}_s + {1\over3\pi}\log(m_t/M_Z)$.}:
\begin{eqnarray}
{1\over\alpha_i^{MSSM}(Q)} &=& {1\over\alpha^{SM}_i(M_Z)} - {C_i\over12\pi}
 + 2 \sum_{k} \Delta b_{ik} \log{M_k \over M_Z}\label{eqn:llt}\\
 &+& 2b^{MSSM}_i \log{Q \over M_Z} +
{\rm ~two-loop~contribution}\nonumber
\end{eqnarray}
where we have made explicit the ~$\overline{MS}\rightarrow\overline{DR}$ ~
conversion factor with ~$C_1=0, ~C_2=2, ~C_3=3$ ~\cite{TAMV}. ~
$M_k$~ are the superpartner masses and ~$\Delta b_{ik}$ ~are their
contributions to the one--loop ~$\beta$ ~functions of the couplings ~
$\alpha_i$. ~
This is the correct result
for the running of the gauge couplings at the two-loop
accuracy as long as the contribution to the SM from the (non-renormalizable)
higher dimension operators, left over after decoupling of superpartners, can
be neglected in the process of extracting $g_i(M_Z)$~ from the data
(we shall call it the  Leading Logarithmic Threshold (LLT)
approximation). This requires ~$M_k\gg M_Z$ ~for all superpartner masses.

Assuming that there are only two GUT scale parameters: ~$\alpha_{U}$ ~
and ~$M_{U}$ ~(i.e. assuming that the potential GUT scale corrections to
the gauge coupling unification are negligible) we can predict one of the
couplings at ~$M_Z$ ~scale, e.g. ~$\alpha_s(M_Z)$ ~in terms of the other two
and of the superpartner masses, which are at present free parameters.
In the LLT approximation
the dependence of the prediction for ~
$\alpha_s(M_Z)$ ~on the supersymmetric spectrum can be described by
a single effective parameter ~$T_{SUSY}$ ~\cite{LP1}:
\begin{eqnarray}
\alpha_s(M_Z) &=& f(\alpha_1(M_Z),\alpha_2(M_Z),T_{SUSY})\\
        &=& {\tilde f}(G_F,M_Z,\alpha_{EM}(M_Z),m_t,M_{\phi^0},T_{SUSY})
\nonumber
\label{eqn:alphapred}
\end{eqnarray}
where \cite{CARENA}:
\begin{eqnarray}
T_{SUSY} = |\mu |
 \left(m_{\tilde W}^2\over m^2_{\tilde g}\right)^{14\over 19}
 \left(M_{A^0}^2\over\mu^2\right)^{3\over38}
 \left(m^2_{\tilde W}\over\mu^2\right)^{2\over 19}
 \prod_{i=1}^3 \left({M_{\tilde L_i}^3 M_{\tilde Q_i}^7 \over
  M_{\tilde E_i}^2 M_{\tilde U_i}^5 M_{\tilde D_i}^3 } \right)^
  {1 \over 19}
\label{eqn:tsusy}
\end{eqnarray}
The effective parametrization in terms of ~$T_{SUSY}$ ~is exact for one--loop
RGE and the correction due to the superpartner spectrum then reads:
\begin{eqnarray}
{1\over\alpha_3^{SM}} = {1\over\alpha_3^0}
+{1\over2\pi}{19\over14}\log{T_{SUSY}\over M_Z}
\label{eqn:corr}
\end{eqnarray}
($\alpha_3^0$ ~is the value predicted without the inclusion of threshold
corrections). With two--loop equations there is some (weak) dependence
on the details of the spectrum through the dependence on the spectrum of
the two--loop contribution on the way up to ~$M_{U}$.

The prediction of the eqns.
(\ref{eqn:llt}--\ref{eqn:corr}) may be subject to important corrections if
some of the superpartner masses are ~${\cal O}(M_Z)$. ~Then the renormalizable
SM is not the correct effective theory at the electroweak scale
and the non--renormalizable terms should be included when extracting
the couplings from the data. Equivalently, we can work at ~$M_Z$ ~in the
framework of the full MSSM, extract from the data the MSSM couplings
including full 1--loop threshold contribution from SUSY loops (not just the
leading logarithms) and study the unification of the MSSM couplings.
(Note that in the LLT approximation an equivalent
interpretation of equation (\ref{eqn:llt}) is:
\begin{eqnarray}
{1\over\alpha_i^{MSSM}(Q)} = {1\over\alpha^{MSSM}_i(M_Z)} &+&
2b^{MSSM}_i \log{Q\over M_Z}
\label{eqn:lltcorr0}
\\
&+&{\rm ~two-loop~contribution} \nonumber
\end{eqnarray}
with
\begin{eqnarray}
{1\over\alpha^{MSSM}_i(M_Z)} = {1\over\alpha^{SM}_i(M_Z)} - {C_i\over12\pi}
+ 2\sum_{k}\Delta b_{ik}\log{M_k\over M_Z}
\label{eqn:lltcorr}
\end{eqnarray}
where RG running with the MSSM ~$\beta-$functions starts directly from ~
$M_Z$, ~and the threshold corrections are absorbed in a redefinition of ~
$g_i(M_Z)^{SM} \rightarrow$\\ $g_i(M_Z)^{MSSM}$.)

The outlined program has been accomplished by several groups:
\cite{unif,barbieri,bagger}. Clearly, the  values of the MSSM couplings
extracted from the data depend now on the superpartner masses
\footnote{Also ~$M_{\phi^0}$ ~must be replaced by ~$M_{A^0}$ ~and ~
$\tan\beta$.} ~$M_k$, e.g.:
\begin{eqnarray}
 \sin^2\theta_W(M_Z)^{MSSM}=f(G_F,M_Z,\alpha_{EM},m_t,M_k,M_{A^0},\tan\beta)
\nonumber
\end{eqnarray}
not only by logarithmic terms as in eq.(\ref{eqn:lltcorr}), but also by
terms ~${\cal O}(M_Z/M_k)$ ~and the additional corrections may be ~
$\sim1$\% ~for ~$\sin^2\theta_W(M_Z)^{MSSM}$~ as shown in ref. \cite{unif}.

In our analysis we also use ~$\alpha_3^{MSSM}$ ~with the oblique
non--logarithmic corrections included \cite{FAGRI,unif} but they are
unimportant for generic spectra which have coloured sparticles rather heavy.
We also use the properly extracted ~$\alpha_{EM}^{MSSM}$ ~\cite{unif}.

The impact of the non-leading SUSY corrections on the prediction
for ~$\alpha_s(M_Z)$ ~is illustrated in Fig. 1 for a generic
sparticle spectrum obtained in the minimal supergravity model
(with universal boundary conditions for the soft SUSY breaking scalar mass
parameters at the GUT scale) with radiative electroweak breaking
and squark masses below 2 TeV \cite{OLPOK}. The results for ~$\alpha_s(M_Z)$ ~
are plotted as a function of ~$T_{SUSY}$ ~defined in eq. (\ref{eqn:tsusy}).
We compare the results obtained in the LLT
approximation for the superpartner thresholds, eq. (\ref{eqn:llt}),
with their complete inclusion at the one-loop level, as in ref. \cite{unif}.
The non-leading corrections increase the predicted value of ~
$\alpha_s(M_Z)$ ~for ~$T_{SUSY} < 100$ GeV. ~We conclude that unification
without GUT threshold corrections predicts e.g. for ~$m_t=160~(180)$ GeV ~
$\alpha_s(M_Z)> 0.126~(0.128)$ ~for ~$T_{SUSY} < M_Z$ ~and ~
$\alpha_s(M_Z)> 0.121~(0.123)$ ~for ~$T_{SUSY} < 300$ GeV. \footnote{
The uncertainty in ~$\alpha_s$~ induced by the errors in eqs.
(\ref{eqn:fits},\ref{eqn:dalfa}) is ~$\sim 0.0015$~ \cite{unif}.}
It is clear from eq.(\ref{eqn:tsusy}) that ~$T_{SUSY}$ ~depends strongly on ~
$\mu,~m_{\tilde W}$, ~$m_{\tilde g}$ ~and weakly on the
other SUSY masses. In models with the GUT relation
\begin{eqnarray}
 M_3(M_{U})=M_2(M_{U})~,
\label{eqn:gutrel}
\end{eqnarray}
to a very good approximation:
\begin{eqnarray}
T_{SUSY} \sim \mu \left( \alpha_2(M_Z) \over \alpha_3(M_Z) \right)^
{3 \over 2} \sim {1 \over 7} \mu
\end{eqnarray}
and large ~$T_{SUSY}$ ~means very large higgsino mass. From the naturalness
of the Higgs potential \cite{BG,RR} it follows then that also the other
sparticle masses are to be heavy. For instance in the generic spectrum
obtained in models with radiative breaking and universal boundary conditions
for the soft scalar masses at the GUT scale ~$T_{SUSY}=300$ GeV ~corresponds
to the squark masses ~${\cal O}$(2 TeV). ~Of course, large values of ~
$T_{SUSY}$ ~can be obtained also for small ~$\mu$ ~with a spectrum which
violates the GUT relation (\ref{eqn:gutrel}), i.e. with a large ratio ~
$m_{\tilde W}/m_{\tilde g}$. ~However, it is very difficult to imagine such
a scenario without losing the motivation for the minimal unification itself.
\vskip 0.5cm
{\bf 3.} The assumption about negligible GUT scale corrections
to coupling unification may be too restrictive. Various groups have
discussed the GUT threshold corrections (dependent on the GUT
model) \cite{JAP,yamada,wright,LP1,LP2} and ~${\cal O}(M_U/M_{Pl})$ ~
corrections \cite{hall,arnath}. Admitting non-negligible but strongly model
dependent  GUT scale corrections means that, strictly speaking, the
predictivity is lost and one can only use the bottom-up approach: measure ~
$g_i(M_Z)$ ~with better and better precision, measure the sparticle spectrum
and study the convergence of the couplings at the large scale. Useful
mismatch parameters then are: ~$D_3(M_{U})$ ~and ~$\Delta_3(M_{U})$ ~
(eqs. (\ref{eqn:di},\ref{eqn:deltai}); ~$M_{U}$ ~is defined by unification
of ~$\alpha_1$ ~and ~$\alpha_2$). ~$\Delta_3(M_{U})$ ~is directly related
to the GUT threshold corrections. Again, neglecting ~${\cal O}(M_Z/M_k)$ ~
non--renormalizable terms, both ~$D_3(M_{U})$ ~and ~$\Delta_3(M_{U})$ ~
are functions of ~$\alpha_i$, ~$i=1,2,3$ ~and the effective ~$T_{SUSY}$. ~
Inclusion of non-leading supersymmetric threshold corrections brings in
additional dependence on the spectrum with, however, ~$T_{SUSY}$~ still a
useful parameter to present the results. We show them as a function of ~
$T_{SUSY}$ ~in the LLT approximation for the SUSY thresholds and with their
complete inclusion in Fig.2 for our generic spectra for $m_t=180$~GeV
, ~$\tan{\beta}=10$ ~and for
 three values of ~$\alpha_s=$0.11, 0.12, 0.13. ~
In Fig.3 we plot the same mismatch parameters as a function of ~$\alpha_s$ ~
for our generic spectra. The LLT results for two fixed values of ~
$T_{SUSY}=300$ GeV ~and 1 TeV are also shown for comparison.

The general conclusion is that in the range ~$\alpha_s(M_Z)=0.11-0.13$ ~and ~
$T_{SUSY}=(20-10^3)$ GeV ~the  gauge couplings do unify within ~
the accuracy better than
7\% for ~$m_t=160$ GeV ~and ~${\cal O} (8\%)$ for ~$m_t=180$ GeV, ~
with the maximal mismatch for low values of ~$\alpha_s(M_Z)$ ~and ~
$T_{SUSY}$. ~Is this  mismatch a lot or a little depends on the
GUT model and the expected magnitude of the GUT scale corrections in it
\cite{JAP,yamada,arnath,bagger}.

\vskip 0.5cm
{\bf 4.} In stringy unification the unification scale is no longer a free
parameter. It is related to the value of the unified coupling
\cite{kaplunovsky}:
\begin{eqnarray}
M_{ST} = g_{ST}\times5.27\times 10^{17}~{\rm GeV}~~
\sim {\cal O}(4\times10^{17}~{\rm GeV})~~
\end{eqnarray}
It is interesting to study within the bottom--up approach the mismatch
parameters $D_3,~D_1, ~\Delta_3,~\Delta_1$ ~
(eqs. (\ref{eqn:di},\ref{eqn:deltai})) at the scale ~
$M_{ST}=4\times10^{17}$ GeV. ~The results as a function of ~$\alpha_s(M_Z)$ ~
are shown in Fig.4. We use again our sample of generic spectra.
The results for very heavy spectra with ~$T_{SUSY}=1$ TeV ~and ~$5$
TeV ~obtained within the LLT approximation are also shown. The general
conclusion which can be drawn from these plots is that the mismatch of the
couplings ~$\alpha_3$ ~and ~$\alpha_2$ ~as well as ~$\alpha_1$ ~and ~
$\alpha_2$ ~at ~$M_{ST}$ ~is ~$>{\cal O}(10\%)$. Therefore, to achieve
unification, the string threshold corrections have to be large at the string
scale and in addition must conspire so that they are small at the GUT scale,
i.e. that the approximate unification occurs at ~
$M_{U}\sim3\times10^{16}$ GeV. ~It is also worth pointing out that the
dependence of ~$D_1(M_{ST})$~ and ~$D_3(M_{ST})$ ~on the supersymmetric
spectrum is different and the spectrum which diminishes the first enhances
the second.

It is possible to take the attitude that the value of ~$\alpha_1$ ~at the
string scale is unconstrained because the Kac--Moody level of the U(1) group
can be treated as a free parameter \cite{ib2} ~$k_1$. \footnote{
In our convention ~$k_1=1$~ in the case of SU(5) -- type unification.
This differs from the definition adopted in ref. \cite{ib2}.}
 ~We then have:
\begin{eqnarray}
k_1 \alpha_1~=~\alpha_2~=~\alpha_3~~~({\rm at}~M_{ST})
\end{eqnarray}
In this case (and for negligible stringy threshold corrections)
our parameter ~$D_1$ ~is related to the parameter ~$k_1$:
\begin{eqnarray}
 k_1 =  {1 \over {D_1+1}}.
\end{eqnarray}
With our generic spectra we get: ~$k_1=0.88~-~0.92$. ~However, even then we
are still faced with a large mismatch between ~$\alpha_3$ ~and ~$\alpha_2$ ~
which for our generic spectra requires large string threshold corrections.

Finally it is interesting to go beyond the discussion based on our generic
spectra and to address the following two questions:

1) Does there exist a pattern of the MSSM  spectrum which shifts the
unification point of all three couplings to $M_{ST}$~with negligible stringy
thresholds?

2) Suppose ~$k_1\ne5/3$ ~(i.e. ~$\alpha_1\ne\alpha_2$ ~at ~$M_{ST}$) ~
and helps to unify $\alpha_1$~and  $\alpha_2$. Are there MSSM spectra
which unify ~$\alpha_3$ ~and ~$\alpha_2$ ~at ~$M_{ST}$~ with negligible
stringy thresholds?

In order to answer these questions it is useful to introduce two new effective
parameters describing the impact of the SUSY spectrum on unification of ~
$\alpha_1$ ~and ~$\alpha_2$ ~and ~$\alpha_2$ ~and ~$\alpha_3$ ~separately.
{}From eqs. (\ref{eqn:lltcorr0}) and (\ref{eqn:lltcorr}) we have:
\begin{eqnarray}
M_{U} = M_{U}^0 \left(M_Z\over T_{SUSY}^\prime\right)^{2\over7}
\label{eqn:m12}
\end{eqnarray}
where ~$M_{U}^0$~~$(M_{U}^0)$~ is the crossing point of ~$\alpha_1$~ and
{}~$\alpha_2$ ~
with SUSY threshold corrections included (neglected) and
\begin{eqnarray}
T_{SUSY}^\prime=\left(M_{A^0}\mu^4m_{\tilde W}^{20}\right)^{1\over25}
{\left(M_{\tilde Q}^7M_{\tilde L}\right)^{1\over8}\over
\left(M_{\tilde U}^4M_{\tilde D}M_{\tilde E}^3\right)^{1\over8}}
\end{eqnarray}
All generations have the same masses in the above formula but
a generalization is straightforward.
Noticing that all (none) of the sparticles in the denominator
(numerator) are SU(2) singlets one can write a simplified formula:
\begin{eqnarray}
T_{SUSY}^\prime= { M_L^2 \over M_R }
\end{eqnarray}
with  obvious definitions of the averages ~$M_L$ ~and ~$M_R$. ~
In Fig.5 we plot ~$M_{U}$ ~resulting from formula (\ref{eqn:m12}). For ~
$M_{U}=4\times10^{17}$ ~we need ~$T_{SUSY}^\prime=2\times10^{-2}$ GeV ~
which means that for ~$M_L=M_Z$ ~we would have ~$M_R=400$ ~TeV. As we can
see the answer to question 1 is negative. Regardless of the ~
$\alpha_2~-~\alpha_3$ ~unification, bringing ~$M_{U}$ ~up to ~$M_{ST}$ ~
would require an unacceptable ~$M_R$. ~

Turning to question 2 we study the correction to ~$\alpha_3^{SM}(M_Z)$ ~
predicted from the condition ~$\alpha_3(M_{ST})=\alpha_2(M_{ST})$, ~induced
by the SUSY thresholds. From (\ref{eqn:lltcorr0}) and \ref{eqn:lltcorr}) we
obtain:
\begin{eqnarray}
{1\over\alpha_3^{SM}}={1\over\alpha_3^0}
+{1\over2\pi}{19\over6}\log{M_D\over M_{S}}
\label{eqn:a3st}
\end{eqnarray}
Where
\begin{eqnarray}
{M_D\over M_S} = \left(M_{A_0}\over M_Z\right)^{1\over19}
\left({m^8_{\tilde W}\mu^4}\over m_{\tilde g}^{12}\right)^{1\over19}
\left({M^3_{\tilde Q} M^3_{\tilde L} }\over{M_{\tilde U}^3
M_{\tilde D}^3}\right)^{1\over19}
\label{eqn:mdms}
\end{eqnarray}
and ~$\alpha_3^0$ ~is the value predicted without the inclusion of SUSY
threshold corrections. In Fig.6 we show ~$\alpha_3$ ~predicted with the use
of formula (\ref{eqn:a3st}) (with ~$\alpha_3^0$ ~obtained from the two--loop
RGE) for ~$M_{ST}=(3.5~,~4.0~,~4.5)\times10^{17}$ GeV ~as a function of the
ratio ~$M_D/M_S$. ~In order to get ~$\alpha_s < 0.13$ ~one needs ~
$M_D/M_S >20$. ~
At this point we disagree with the recent analysis of ref.
\cite{kobayashi} which reconciles ~$\alpha_s(M_Z)=0.118$ ~with
the stringy unification for the SUSY spectra with smaller hierarchies.
Taking masses in the numerator of our eq. (\ref{eqn:mdms}) ~
$\sim 30$ ~times larger than masses in the denominator we still get ~
$\alpha_s \geq 0.125$ ~as can be seen from Fig.6
\footnote{For ~$\alpha_s(M_Z)\simeq0.12$ ~the masses should be split by
a factor of at least 60.}.
This disagreement is mainly due to the use of one--loop RGE in ref.
\cite{kobayashi} and somewhat higher value of ~$\sin^2 \theta_W(M_Z)$ ~
(more appropriate for ~$m_t \sim 160$ GeV). ~The ratio ~$M_D/M_S$ ~is
dominated by the ratio ~$m^2_{\tilde W}\mu/m_{\tilde g}^3$ ~which in the
case of string unification
is more model dependent than for GUTs \cite{munoz,kapluis}. In particular
it is conceivable in this case that ~$m_{\tilde W}>m_{\tilde g}$. ~ For ~
$M_S=M_Z$ ~we would get ~$M_D>2$ TeV ~but in fact for so light spectrum the
non--logarithmic corrections could raise the predicted value of ~$\alpha_s$ ~
as is evident from Fig.1. For ~$M_S=150$ GeV, ~when non--logarithmic effects
are small, we get ~$M_D>3$ TeV. ~We conclude that it is possible to raise the ~
$\alpha_2 - \alpha_3$ ~unification scale up to ~$M_{ST}$ ~but only with
highly unnatural SUSY spectra,
with the heaviest sparticles above 3 TeV
and with ~$\alpha_s(M_Z)\simeq 0.13$. ~
Otherwise large string threshold corrections are needed.
\vskip 0.5cm
{\bf 5.} We have discussed the impact of SUSY thresholds on the unification of
gauge couplings in the framework of GUT and string theories.
Non--logarithmic SUSY corrections can be important for the phenomenologically
interesting case of light superpartners. These corrections always reduce the
value of ~$\sin^2\theta_W(M_Z)^{MSSM}$ ~which in turn raises the value of ~
$\alpha_s(M_Z)^{SM}$ ~predicted from SUSY unification. In the minimal
unification scenario (i.e. with negligible GUT scale corrections to the
running of the couplings) one gets ~$\alpha_s(M_Z) >0.121 (123)$ ~for the
effective  parameter ~$T_{SUSY}<300$ GeV ~and  ~$\alpha_s(M_Z)>0.115 (117)$ ~
for ~$T_{SUSY}<1$ TeV, ~for ~$m_t=$160 ~and ~180 GeV, ~respectively.
For the generic spectra in the minimal supergravity model ~$T_{SUSY}\sim1$
TeV ~corresponds to very heavy sfermions, e.g. squark masses are ~
${\cal O}(5$~TeV). ~More generally in the bottom--up running the couplings
do unify within a few percent accuracy even for low ~$\alpha_s(M_Z)$ ~and small
values of ~$T_{SUSY}$, ~e.g. the mismatch between ~$\alpha_3$ ~and ~
$\alpha_2$ ~at the scale of unification of ~$\alpha_1$ ~and ~$\alpha_2$ ~is
generically below ~${\cal O}(5\%)$. ~The mismatch of the couplings at ~
$M_{ST}=4\times10^{17}$ GeV ~is much larger, typically ~${\cal O}(10\%)$ ~or
more, and it cannot be eliminated by any sensible  superpartner  spectrum.
String unification requires, therefore, large string threshold corrections
(which, however, may not be unrealistic \cite{nilles}) which  conspire
to give the effective unification scale ~$\sim3\times10^{16}$ GeV. ~The
scenario
with ~$\alpha_1$ ~and ~$\alpha_2$ ~unified by treating the Kac--Moody level ~
$k_1$ ~as a free parameter is not particularly helpful with regard to
the coupling unification at ~$M_{ST}$ ~(and is rather uneconomical).

\newpage
\noindent {\bf FIGURE CAPTIONS}
\vskip 0.5cm

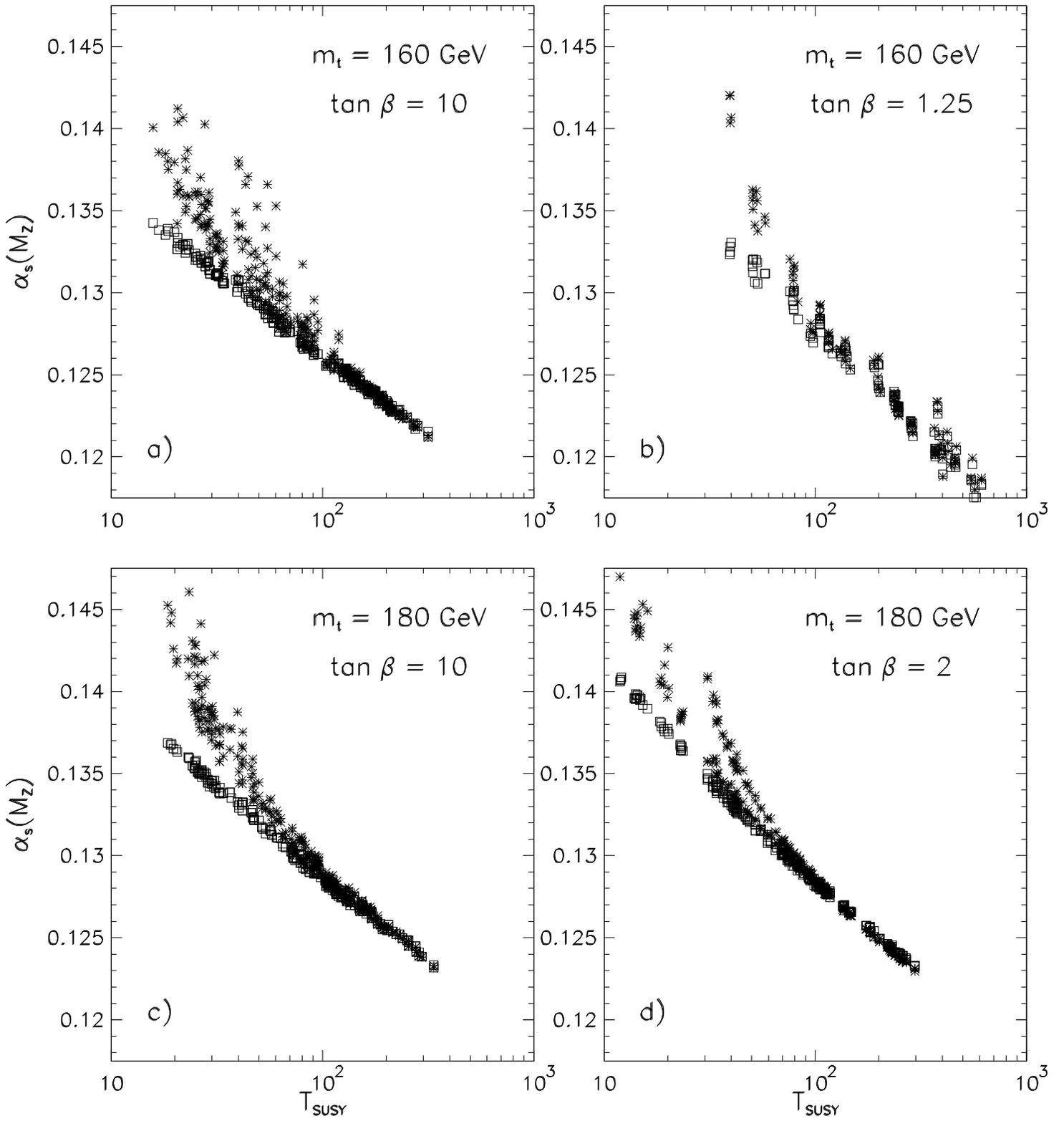
\noindent {\bf Figure 1.}

\noindent
$\alpha_s(M_Z)$ ~predicted by the minimal unification
as a function of ~$T_{SUSY}$ ~for ~different values of ~$m_t$ ~and ~
$\tan\beta$. ~Squares (stars) correspond to the LLT (full) calculation
of the supersymmetric thresholds for a generic sample of SUSY spectra
obtained in the minimal supergravity model with radiative electroweak
breaking and universal boundary conditions.
\vskip 0.3cm

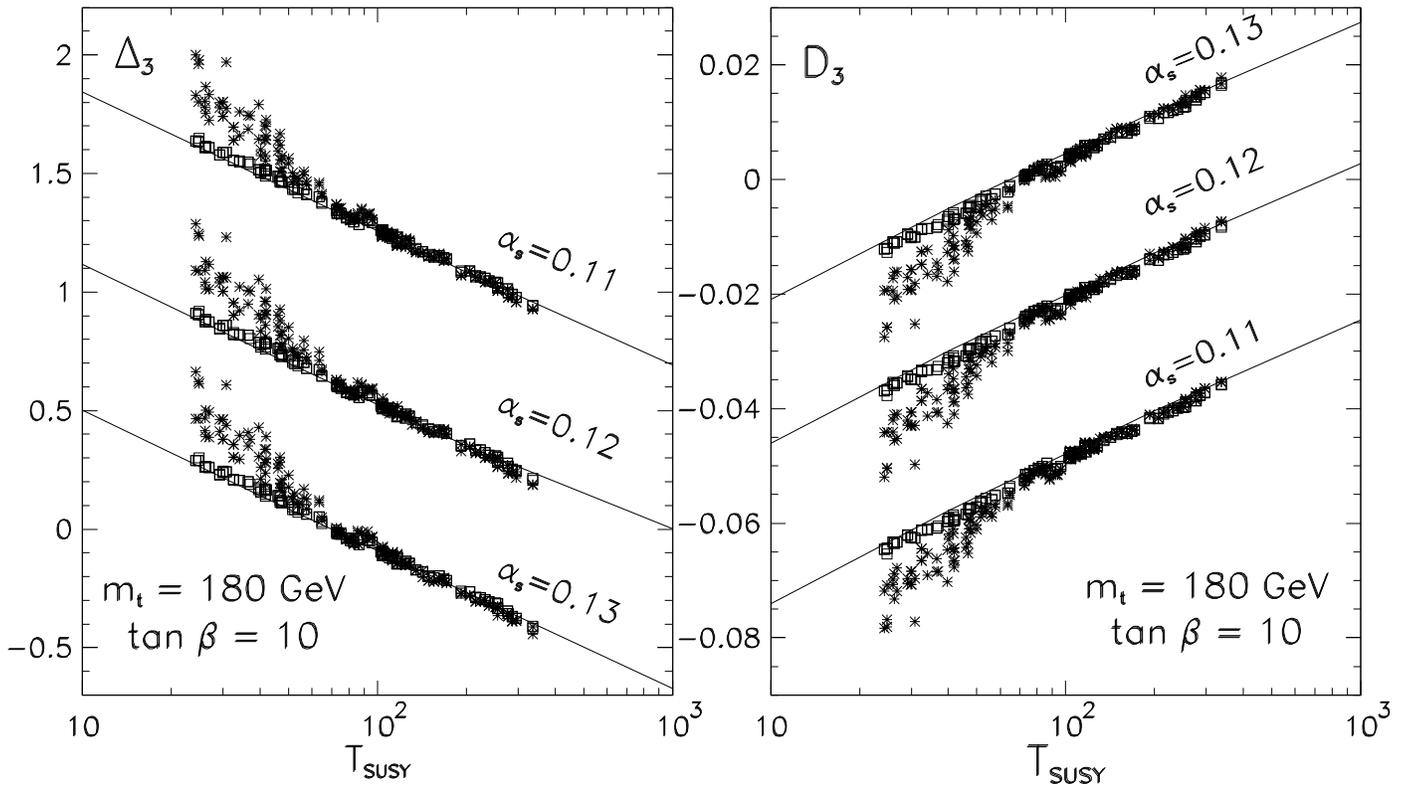
\noindent {\bf Figure 2.}

\noindent
Mismatch parameters ~$\Delta_3(M_{U})$ ~and ~$D_3(M_{U})$ ~for ~$m_t=180$ ~
and ~$\tan\beta=10$ ~as a function of ~$T_{SUSY}$ ~for several values of ~
$\alpha_s$. ~Squares (stars) correspond to the LLT (full) calculation of the
supersymmetric thresholds. Sample of spectra as in Fig.1c. Solid lines
extrapolate ~$\Delta_3(M_{U})$ ~and ~$D_3(M_{U})$ ~
in the LLT approximation up to ~$T_{SUSY}=1$
TeV.
\vskip 0.3cm

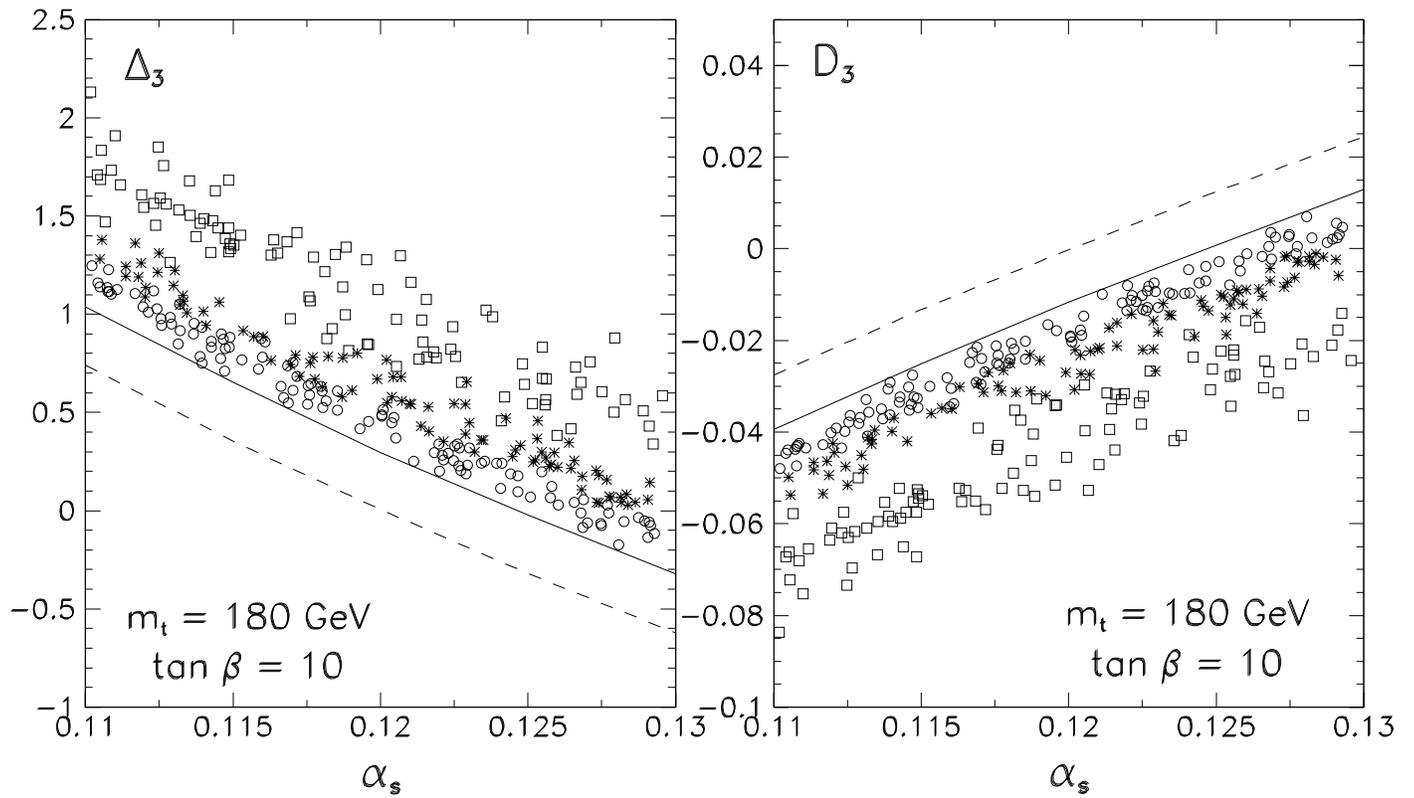
\noindent {\bf Figure 3.}
\noindent
Mismatch parameters ~$\Delta_3(M_{U})$ ~and ~$D_3(M_{U})$ ~as a function
of ~$\alpha_s$ ~for  the same sample of
spectra as in Fig.1c. Squares, stars and circles show the results of the
full calculation for spectra with ~$M_{\tilde Q}<500$ GeV, ~
$500~{\rm GeV}<M_{\tilde Q}<1$ TeV ~and $1~{\rm TeV}<M_{\tilde Q}<2$ TeV.
respectively. For comparison the LLT calculation for a spectra with ~
$T_{SUSY}=$300 GeV ~(1 TeV) ~are marked by the solid (dashed) lines.
\vskip 0.3cm

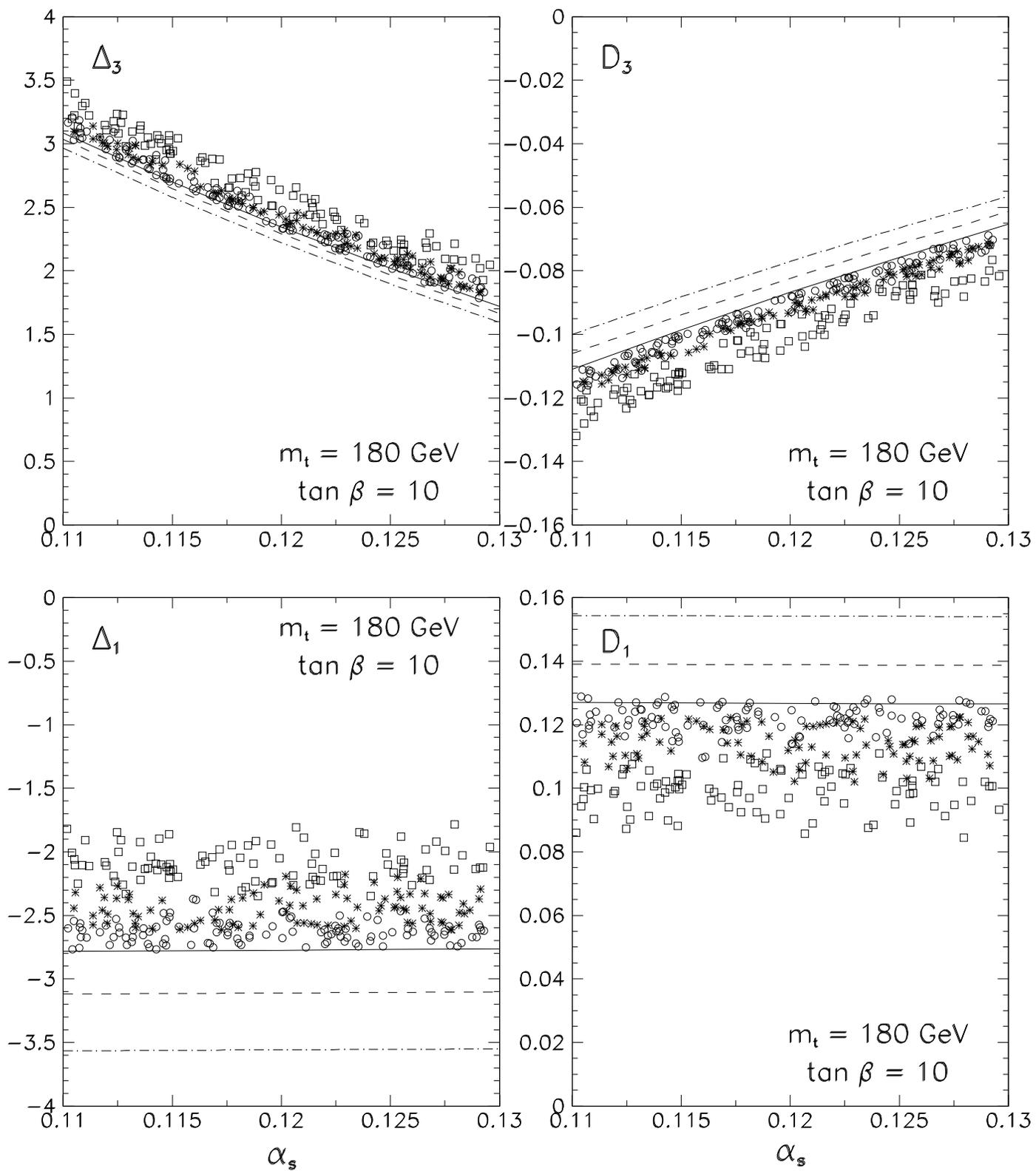
\noindent {\bf Figure 4.}

\noindent
Mismatch parameters ~$\Delta_3$, ~$D_3$, ~$\Delta_1$ ~and ~$D_1$ ~at the
string scale ~$M_{ST}=4\times10^{17}$ ~as a function of ~$\alpha_s$ ~
for our generic spectra. Markers as in Fig.3. Solid, dashed and
dash--dotted
lines correspond to the LLT calculation for
$T_{SUSY}=$300 GeV, ~1 ~and ~5 TeV ~respectively.
\vskip 0.3cm

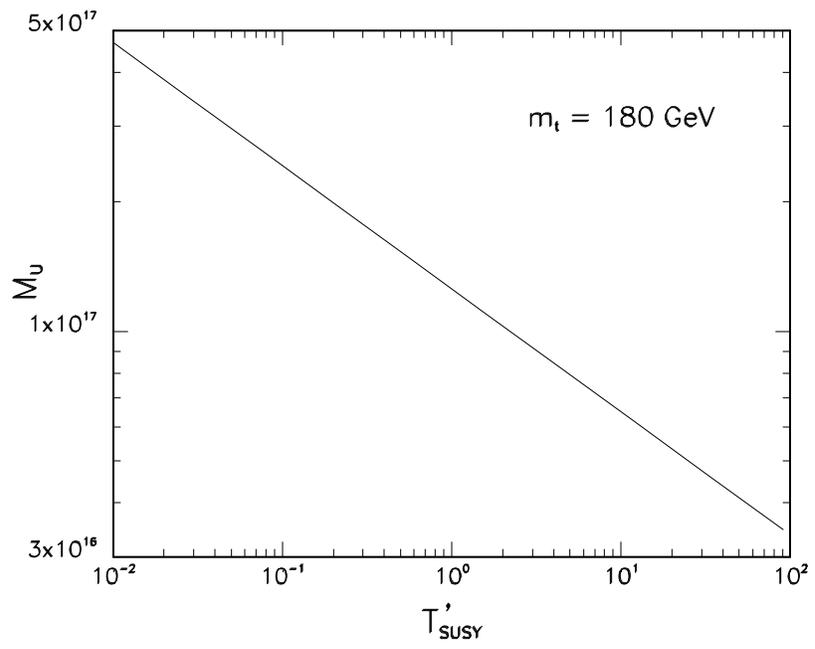
\noindent {\bf Figure 5.}
$M_{U}$ ~as a function of ~$T_{SUSY}^{\prime}$ ~from eq. (\ref{eqn:m12})
for ~$m_t=180$ ~GeV.
\noindent

\vskip 0.3cm

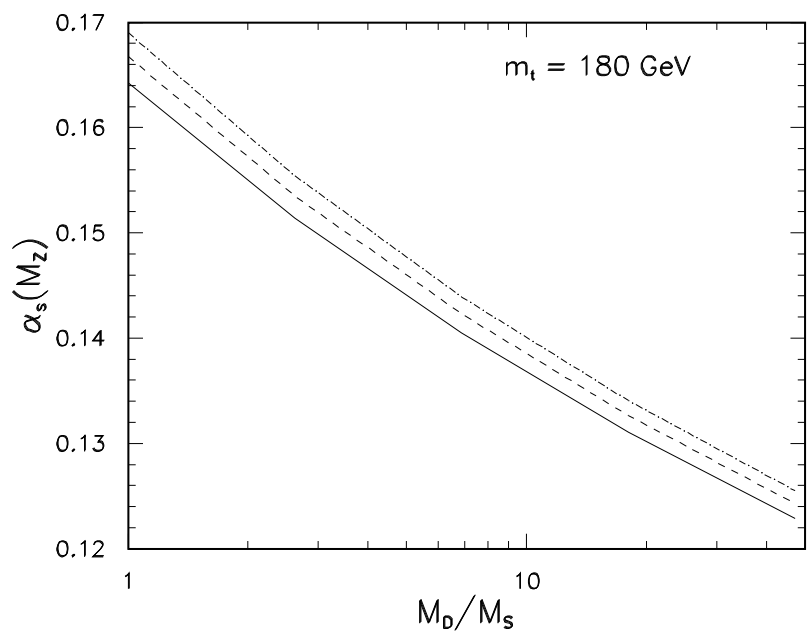
\noindent {\bf Figure 6.}

\noindent
Prediction for ~$\alpha_s$ ~from the condition ~
$\alpha_3(M_{ST})=\alpha_2(M_{ST})$ ~with ~$M_{ST}=3.5$, ~(solid) ~4.0 ~
(dashed) ~and ~$4.5\times10^{17}$ GeV ~(dash--dotted) ~with the use of eq.
(\ref{eqn:a3st}) as a function of the parameter ~$M_D/M_S$ ~for ~$m_t=180$ GeV.
\vskip 0.3cm
\end{document}